\documentclass[aps,prd,reprint,superscriptaddress,nofootinbib,10 pt]{revtex4}

\usepackage{amsmath,amssymb,amsfonts}
\usepackage{graphicx}
\usepackage{xcolor}
\usepackage{hyperref}
\usepackage{braket}
\usepackage{booktabs}
\usepackage{multirow}
\usepackage{cases}
\newcommand{\bse}{\begin{subequations}}
\newcommand{\ese}{\end{subequations}}
\newcommand{\be}{\begin{equation}}
\newcommand{\ee}{\end{equation}}
\newcommand{\bea}{\begin{eqnarray}}
\newcommand{\eea}{\end{eqnarray}}
\newcommand{\ba}{\begin{array}}
\newcommand{\ea}{\end{array}}

\makeatletter
\renewcommand{\eqref}[1]{\textcolor{blue}{(\ref{#1})}}
\makeatother

\hypersetup{
    colorlinks=true,
    linkcolor=blue,
    citecolor=green,
    urlcolor=cyan,
    filecolor=magenta
}

\makeatletter
\def\tagform@#1{\maketag@@@{\textcolor{blue}{(\ignorespaces#1\unskip\@@italiccorr)}}}
\makeatother

\begin{document}
\title{Negative and Positive Saddle-Selection Ambiguities in the HTEE}
\author{H. Ebrahim}
\email{hebrahim@ut.ac.ir}
\affiliation{Department of Physics, University of Tehran, North Karegar Avenue, Tehran 14395-547, Iran}
\author{M. Ali-Akbari}
\email{m\_aliakbari@sbu.ac.ir}
\author{K. Taheranvarian}
\email{k.taheranvarian@mail.sbu.ac.ir}
\affiliation{Department of Physics, Shahid Beheshti University, 1983969411, Tehran, Iran}

\begin{abstract}
The holographic timelike entanglement entropy (HTEE), computed via complexified extremal surfaces, suffers from a saddle-selection ambiguity: for sufficiently small time intervals (\(\Delta t\to0\)), both a vacuum-connected and a vacuum-disconnected complex extremal surface contribute, and the minimal-real-part criterion and the vacuum-reduction criterion can favor different saddles. We study this ambiguity in strongly coupled field theories whose dual geometries are hyperscaling-violating Lifshitz backgrounds, in which conformal symmetry and Lorentz invariance are broken by the Lifshitz exponent and the hyperscaling-violation parameter. Working in \(2+1\)- and \(3+1\)-dimensional dual field theories, we derive general expressions for the HTEE in the vacuum-connected and vacuum-disconnected saddles, \(s_{v.c.}\) and \(s_{v.d.}\), respectively, and compare their real parts. We find that the two criteria agree everywhere except at the first accidental zero. The ambiguity appears in two distinct forms. In the \emph{negative} family, \(\operatorname{Re}(s_{v.c.})=0\) and \(\operatorname{Re}(s_{v.d.})<0\), whereas in the \emph{positive} family, \(\operatorname{Re}(s_{v.c.})>0\) and \(\operatorname{Re}(s_{v.d.})=0\). The latter shows that the ambiguity survives even in the conformal \(3+1\)-dimensional case. It therefore seems that the saddle-selection ambiguity is a generic feature of the complexified-geometry prescription.
\end{abstract}

\maketitle

\tableofcontents

\section{Introduction}
\noindent
The standard perturbation method cannot be applied to strongly coupled field theories because the coupling constant is not sufficiently small. Therefore, alternative nonperturbative methods have been proposed and applied, such as gauge-gravity duality \cite{Maldacena:1997re,Aharony:1999ti,Nastase:2007kj,Casalderrey-Solana:2011dxg,Natsuume:2014sfa,Baggioli:2019rrs}. This duality states that different physical processes or quantities in a strongly coupled gauge theory can be described by their adequate counterparts in a weakly coupled gravity theory in a higher dimension. Over more than two decades, various properties of the duality have been studied, revealing that the gravity counterparts generally have a simple interpretation. This has opened a gate to a deeper understanding of the virtues of strongly coupled systems. Utilizing the duality, one significant class of quantities that has attracted a lot of attention is entanglement measures, such as spacelike entanglement entropy \cite{Ryu:2006bv,Nishioka:2009un}, timelike entanglement entropy \cite{Doi:2023zaf}, mutual information \cite{Hayden:2011ag,Headrick:2019eth}, and entanglement of purification \cite{Takayanagi:2017knl}. In particular, holographic entanglement measures provide powerful geometric probes of strongly coupled dynamics that are often difficult to access by conventional field-theoretic techniques, and they continue to offer new insights into the structure of quantum entanglement in strongly interacting systems.

To define the entanglement entropy \cite{Calabrese:2004eu,Casini:2009sr}, let us consider a pure state \(\ket{\phi}\) with density matrix \(\rho=\ket{\phi}\bra{\phi}\), and suppose that the total system is spatially divided into two complementary subsystems, \(A\) and \(\bar{A}\). By tracing over the degrees of freedom of the complementary subsystem \(\bar{A}\), one obtains the reduced density matrix
$\rho_A = \mathrm{Tr}_{\bar{A}}\,\rho$.
The entanglement entropy is then defined as the von Neumann entropy of this reduced density matrix, $E_A = -\mathrm{Tr}_A\!\left(\rho_A \log \rho_A\right)$.
This quantity measures the amount of quantum information shared between the two subsystems and, equivalently, the information that becomes inaccessible when one of the subsystems is traced out. For a pure state, it follows immediately that \(E_A=E_{\bar{A}}\). The entanglement entropy has various properties, such as non-negativity, vanishing if and only if the state is a product state with respect to the bipartition, subadditivity and strong subadditivity \cite{Headrick:2007km}. In quantum field theory, \(E_A\) is typically ultraviolet divergent and requires regularization; nevertheless, in many ground states of gauge theories it is expected to obey an area law.

According to the Ryu--Takayanagi (RT) proposal \cite{Ryu:2006bv,Ryu:2006ef}, in strongly coupled gauge theories admitting a holographic dual, the holographic entanglement entropy of a subregion \(A\) is obtained by finding the minimal-area bulk surface \({\cal A}^R\) that is anchored to the boundary of \(A\), i.e.\ \(\partial {\cal A}^R = \partial A\), on the asymptotic boundary where the dual field theory lives. More precisely, the holographic entanglement entropy is given by
\begin{equation}
E_A = \frac{\mathrm{Area}({\cal A}^R)}{4G_N},
\end{equation}
where \(G_N\) is the Newton constant in the dual gravity theory and \({\cal A}^R\) is the real minimal homologous surface in the bulk whose boundary coincides with \(\partial A\). This geometric prescription provides a powerful holographic bridge between quantum entanglement in the boundary field theory and the classical geometry of the higher-dimensional bulk background, and it has been generalized to include quantum corrections, higher-derivative gravity, and time-dependent settings \cite{Faulkner:2013ana,Dong:2013qoa,Hubeny:2007xt}.

A natural generalization of spacelike entanglement entropy to timelike entanglement entropy has been proposed recently. The two notions are conceptually similar: entanglement entropy is defined with respect to a bipartition of the state into subsystems, but for timelike entanglement entropy the separating subregion is timelike rather than spacelike. Thus, timelike entanglement entropy characterizes the degree of entanglement between two subsystems when the entangling subregion is temporal. Its holographic dual for two-dimensional conformal field theories was introduced in \cite{Doi:2022iyj,Doi:2023zaf} by analytic continuation of the spacelike holographic entanglement entropy. Unlike the standard holographic entanglement entropy, this analytic continuation generically leads to an imaginary part for the HTEE. Another method for obtaining the HTEE in arbitrary dimensions was then introduced as a piecewise geodesic construction involving both spacelike and timelike segments \cite{Doi:2023zaf}. The idea of a complexified bulk geometry was subsequently introduced in \cite{Heller:2024whi}, and the same two-dimensional HTEE was computed within this framework. In this approach, the HTEE is given by
\begin{equation}
S = \frac{\mathrm{Area}(\mathcal{A})}{4G_N},
\label{eq:HTEE}
\end{equation}
where the extremal area \(\mathcal{A}\) is obtained in a complexified geometry. These approaches have been applied to investigate various properties of the HTEE; for instance, see \cite{Li:2022tsv,Ali-Akbari:2026oez,Grieninger:2023knz}.
Although the idea of a complexified bulk geometry provides a geometric counterpart for the HTEE and can be applied to different bulk geometries dual to different gauge theories, this prescription suffers from two ambiguities \cite{Heller:2024whi}:
\begin{enumerate}
\item The turning point of the complex extremal surface is typically of the form \(z_*=\pm i|z_*|\). The choice of sign flips the sign of the imaginary part of the HTEE while leaving the real part unchanged, and there is no intrinsic criterion for identifying which sign corresponds to the correct physical result.

\item For sufficiently small time intervals, \(\Delta t\to 0\), two
families of solutions are known to exist: the vacuum-connected
(\(v.c.\)) solutions, characterized by \(z_*\to 0\), and the
vacuum-disconnected (\(v.d.\)) solutions, identified by
\(z_*\to\infty\), both corresponding to the same time interval
\(\Delta t\). It turns out that the \(v.d.\) branch has a smaller real part than the \(v.c.\) branch, while the \(v.c.\) solution reduces to the vacuum answer. Consequently, it is not obvious which criterion determines the physical HTEE: the \(v.d.\) solution with the minimal real part, or the solution that reduces to the vacuum answer.
\end{enumerate}

In this paper, we study holographically the second ambiguity in various non-conformal field theories whose dual geometries are hyperscaling-violating Lifshitz spacetimes. These theories break conformal symmetry and Lorentz invariance, with the amount of breaking controlled by the Lifshitz exponent \(\xi\) and the hyperscaling-violation parameter \(\theta\). The ambiguity we address is the one between the $v.c.$ and $v.d.$ extremal surfaces that contribute to the HTEE for the same boundary subregion. We determine the conditions under which this ambiguity arises as a function of \(\xi\) and \(\theta\), and compare the on-shell actions of the two saddles. We analyze this in detail for \(d=2\) and \(d=3\) dual field theories and identify the parameter ranges in which the $v.c.$ or $v.d.$ solution is favored.

\section{Setup: HTEE and Complex Extremal Surfaces in Hyperscaling-Violating Lifshitz Geometry}
According to gauge--gravity duality, the hyperscaling-violating Lifshitz geometry is holographically dual to a strongly coupled quantum field theory that is neither relativistic nor conformal \cite{Fonda:2014ula}. The dual field theory exhibits anisotropic scaling between time and space, governed by a Lifshitz exponent \(\xi\), which reflects the breaking of Lorentz invariance. The hyperscaling-violation parameter \(\theta\) modifies the effective dimension of the dual field theory, and thermodynamic quantities scale with an effective spatial dimension \(D=d-\theta\), where \(d+1\) is the spacetime dimension in which the field theory lives. Depending on the values of \(\xi\) and \(\theta\), the asymptotic geometry can be AdS, Lifshitz, or hyperscaling-violating Lifshitz. More precisely, AdS and Lifshitz arise as special limits when the hyperscaling-violation parameter vanishes, AdS when the Lifshitz exponent equals one, and Lifshitz when it differs from one, while a nonzero hyperscaling-violation parameter generally gives hyperscaling-violating Lifshitz behavior rather than pure AdS or pure Lifshitz.

The simplest \((d+2)\)-dimensional geometry incorporating Lifshitz scaling and the hyperscaling-violation exponent is given by \cite{Fonda:2014ula}
\be\label{metric}
ds^2=z^{-2D/d}\left(-z^{2(1-\xi)}f\,dt^2+\frac{dz^2}{f}+d\vec{x}^{\,2}\right),
\ee
where \(\vec{x}=(x_1,\dots,x_d)\) and the blackening factor is
\be
f(z)=1-\left(\frac{z}{z_h}\right)^{D+\xi}.
\ee
Here \(z\) denotes the radial coordinate, and the boundary of the geometry is located at \(z\to 0\). For a consistent gravity solution, we impose the following conditions:
\bse\label{conditions}
\begin{align}
\label{c1} D+\xi&> 0,\\
\label{c2} D\left(\xi-1-\frac{\theta}{d}\right)&\ge 0,\\
\label{c3} (\xi-1)(D+\xi)&\ge 0.
\end{align}
\ese
Requiring \(D+\xi>0\), i.e. a positive exponent in the blackening factor, together with \eqref{c3}, forces \(\xi\ge 1\). However, \(\theta\) is not fixed by this requirement alone; it remains subject to the remaining consistency conditions.
Finally, the temperature of the background, identified with that of the dual field theory, is given by
\be
T=\frac{D+\xi}{4\pi z_h^{\xi}}
      =\frac{d-\theta+\xi}{4\pi z_h^{\xi}}.
\ee
The metric \eqref{metric} reduces to \(\mathrm{AdS}_{d+2}\) in Poincaré coordinates when \(\xi=1\) and \(\theta=0\). For further details, we refer the interested reader to \cite{Fonda:2014ula,Dong:2012se}.

As usual, to compute the HTEE, we introduce the subregion
\be\label{interval}
A=\left\{|t|\le\frac{\Delta t}{2},\quad x_1=0,\quad |x_i|\le\frac{L}{2}\ \ (i=2,\dots,d-1)\right\},
\ee
where \(L\to\infty\) and \(\Delta t\) is the length of the time interval. Using the symmetries of the geometry, the complexified extremal surface corresponding to the HTEE can be described by \(t(z)\) and is obtained from the area functional
\be\label{area}
\frac{\mathcal{A}}{2}=V_{d-1}\int dz\ z^{-D}\sqrt{\frac{1}{f}-z^{2(1-\xi)}f\dot{t}^{\,2}}\equiv V_{d-1}\int dz\ \mathcal{L},
\ee
where \(V_{d-1}=\int dx_2\cdots dx_d\) is the volume of the transverse directions. The extremal surface is fully specified by the boundary condition \(t(\epsilon)=\pm\Delta t/2\), where \(\epsilon\) is a UV cutoff. Since \eqref{area} does not depend explicitly on \(t\), the coordinate \(t\) is cyclic, and hence there is a conserved momentum \(p\):
\be
p=\frac{\partial\mathcal{L}}{\partial\dot{t}}.
\ee
Consequently, using the above equation, we immediately obtain
\be\label{dot}
\dot{t}=\frac{p}{fz^{1-\xi}\sqrt{p^2+z^{2\alpha}f}},
\ee
where \(\alpha=1-\xi-D\). For the extremal surface to be smooth at the turning point, we impose
\be
p^2+z_*^{2\alpha}f(z_*)=0,
\ee
where \(z_*\) is the turning point of the extremal surface, i.e.\ the deepest point reached by the surface; at this point \(\dot{t}(z_*)\to\infty\). Substituting \eqref{dot} into \eqref{area}, we obtain
\be
\label{area2}
\frac{\mathcal{A}}{2}=V_{d-1}\int_{z_*}^{\epsilon}dz\ \frac{z^{-D+\alpha}}{\sqrt{p^2+z^{2\alpha}f}},
\ee
and integrating \eqref{dot} gives
\be
\label{time}
\frac{\Delta t}{2}=\int_{z_*}^{\epsilon}dz\ \dot{t}(z).
\ee
The relation between \(\Delta t\) and the turning point \(z_*\) is generally given by \eqref{time}. Substituting this relation into \eqref{area2}, one obtains the extremal surface in terms of the time interval. Using \eqref{eq:HTEE}, the HTEE corresponding to the subregion \eqref{interval} is then computed. This method has been applied to calculate the HTEE in various backgrounds \cite{Ali-Akbari:2026xzx,Afrasiar:2024ldn}.

Since we wish to discuss further the second ambiguity raised in the Introduction, we now consider two families of solutions first presented in \cite{Heller:2024whi}:
\bse\label{alphacondition}
\begin{align}
\label{vc}
v.c. :&\quad z_*\to 0,\quad |p|\to\infty,\quad \text{and}\quad \dot{t}\to\infty,\\
\label{vd}
v.d. :&\quad z_*\to\infty,\quad |p|\to 0,\quad \text{and}\quad \dot{t}\to 0,
\end{align}
\ese
with the same boundary condition and $\alpha<-1$. Other regions of $\alpha$ is discussed in appendix \ref{A1}. These are distinct solutions of the same equations and therefore contribute separately to the HTEE. In the $v.d.$ case, \(z_*\gg z_h\), so the extremal surface extends into the black-hole interior, behind the horizon, whereas in the $v.c.$ case the turning point lies near the boundary, \(z_*\ll z_h\). In both cases, the small-time-interval limit \(T\Delta t\ll 1\) is relevant. In this regime, the $v.d.$ ($v.c.$) solution does (does not) penetrate deeply into the bulk. The relative contribution of the two solutions to the HTEE is then determined by comparing their on-shell actions \eqref{area2}.

Using the $v.d.$ limit \eqref{vd} in the area functional \eqref{area2}, the area of the complexified extremal surface becomes
\begin{equation}
\frac{\mathcal{A}_{v.d.}}{2}
=V_{d-1}\frac{z_h^{1-D}}{D+\xi}
\left[
B\left(\frac{1-D}{D+\xi},\frac{1}{2}\right)
+iB\left(\frac{1}{2}-\frac{1-D}{D+\xi},\frac{1}{2}\right)
\right],
\end{equation}
where \(B(a,b)\) denotes the Euler beta function. Similarly, using the $v.c.$ limit \eqref{vc} in \eqref{area2}, the extremal-surface area is
\begin{equation}\label{Avc}
\frac{\mathcal{A}_{v.c.}}{2}
=-V_{d-1}\frac{z_*^{\,1-D}}{2\alpha}\,
B\!\left(\frac{D-1}{2\alpha},\frac12\right).
\end{equation}
For the $v.c.$ solution, \eqref{time} gives
\begin{equation}\label{tvc}
\Delta t=-\frac{i\,z_{*}^{\xi}}{\alpha}\,
B\!\left(\frac{1}{2}-\frac{\xi}{2\alpha},\,\frac{1}{2}\right).
\end{equation}
Eliminating the turning point \(z_*\) between \eqref{Avc} and \eqref{tvc} in favor of the time interval \(\Delta t\), and applying the gauge-gravity dictionary for the HTEE \eqref{eq:HTEE}, we finally obtain
\begin{equation}\label{svc}
s_{v.c.}
\equiv \frac{4G^{d+2}_N}{V_{d-1}}S_{v.c.}
=\frac{\mathcal{A}_{v.c.}}{V_{d-1}}
=-\frac{1}{\alpha}\left[\frac{i\alpha\,\Delta t}
{B\!\left(\frac12-\frac{\xi}{2\alpha},\frac12\right)}
\right]^{(1-D)/\xi}
B\!\left(\frac{D-1}{2\alpha},\frac12\right).
\end{equation}
Similarly, for the $v.d.$ solution, we find
\begin{equation}\label{svd}
s_{v.d.}
\equiv\frac{4G^{d+2}_N}{V_{d-1}}S_{v.d.}
=\frac{\mathcal{A}_{v.d.}}{V_{d-1}}
=\frac{2z_h^{1-D}}{D+\xi}
\left[
B\left(\frac{1-D}{D+\xi},\frac{1}{2}\right)
+iB\left(\frac{1}{2}-\frac{1-D}{D+\xi},\frac{1}{2}\right)
\right].
\end{equation}
Note that \(s_{v.d.}\) is independent of \(\Delta t\); hence, in this limit, the \(v.d.\) saddle contributes a \(\Delta t\)-independent term to the HTEE, while the \(\Delta t\)-dependence is carried by the \(v.c.\) saddle in \eqref{svc}. Moreover, we would like to mention that, as done in \cite{Heller:2024whi}, the regular part of the extremal surface, i.e. \({\cal A}\equiv {\cal A}_{\mathrm{reg}}\), is considered and discussed throughout this paper.

As a crosscheck, we consider the well-known case for which $d=2$, $\xi=1$ and $\theta=0$ leading to AdS$_4$ geometry which is dual to $2+1$ dimentional conformal field theory. For $v.c.$ solution, it is then easy to see that 
\be\label{vc3}
s_{v.c.}
=
-\frac{i}{4\Delta t}\,
B\!\left(\frac34,\frac12\right)^{2},
\ee
which the HTEE is purely imaginary and thus $\operatorname{Re}(s_{v.c.})=0$. For the $v.d.$ solution we have 
\be\label{vd3}
z_h s_{v.d.}=-\frac{1}{3}B\!\left(\frac23,\frac12\right)
+\frac{2i}{3}B\!\left(\frac56,\frac12\right),
\ee
where
\be
z_h\operatorname{Re}(s_{v.d.})=-\frac13
B\!\left(\frac23,\frac12\right).
\ee
Using the relations between beta and gamma functions,
one can show that the \eqref{vc3} and \eqref{vd3} reproduce the ones reported in \cite{Heller:2024whi}.

The point we would like to emphasize here is that, since the real part of the \(v.c.\) solution vanishes, we have 
\begin{equation}\label{condition}
z_h\,\mathrm{Re}(s_{v.c.}) > z_h\,\mathrm{Re}(s_{v.d.}),
\end{equation}
where the minimal-real-part condition selects the \(v.d.\) solution. This condition is the standard criterion used to identify the physically admissible extremal surface in the holographic entanglement entropy. On the other hand, the \(v.c.\) solution smoothly reduces to the vacuum answer, in the same way as in holographic entanglement entropy. This leads to an ambiguity as to which of the two solutions is physical: the one with the minimal real part, or the one that reduces to the vacuum solution. If the \(v.d.\) solution is selected, then the UV-IR correspondence is violated by the HTEE, and a UV-UV character appears instead. For the \(v.c.\) solution, however, the standard UV-IR correspondence remains valid. Thus, the two criteria, the minimal-real-part condition and the vacuum-limit criterion, favor different saddles, and it seems that an additional physical input is required to resolve which saddle dominates the HTEE. 

\section{Vacuum-Connected and disconnected HTEE in 2+1 field theories}
In this section, we begin with $d=2$, where the geometry reduces to an AdS$_4$ background for $\xi=1$ and $\theta=0$, and we discuss the $v.c.$ and $v.d.$ solutions in the presence of hyperscaling violation and a Lifshitz exponent. In particular, we compare the real parts of these two families of solutions in order to determine which of them is physically favored. Based on our results will be obtained, in last subsection, a more general case is investigated. 
\\
\\
\subsection{2+1-Dimensional Case with $\xi=1$: General $\theta$ and Accidental Zeros}
Let us start with \(d=2\), Lifshitz exponent \(\xi=1\), and general \(\theta\). We then have \(D=2-\theta\) and \(\alpha=\theta-2\). Substituting \(d\), \(\xi\), and \(D\) into the consistency conditions \eqref{conditions} gives
\bse \label{newcon1}
\begin{align}
&\label{conda1} \eqref{c1}:\
\theta < 3, \\
&\label{condb1} \eqref{c2}:\
\theta\le 0\ \text{ or }\ \theta> 2,\\
&\eqref{c3}:\ \text{is automatically satisfied}.
\end{align}
\ese
Intersecting \eqref{conda1} with \eqref{condb1} and by considering $\alpha<-1$, the admissible range is 
\begin{equation}
\theta\in(-\infty,0]-
\end{equation}
Simplifying \eqref{svc} gives
\be\label{svc0}
{\rm Re}(s_{v.c.})=
\frac{B\!\left(\frac{\theta-1}{2(2-\theta)},\frac12\right)}{2-\theta}\left[\frac{(2-\theta)\Delta t}
{B\!\left(\frac{3-\theta}{2(2-\theta)},\frac12\right)}
\right]^{\theta-1}\cos\!\left(\frac{\pi(\theta-1)}{2}\right). 
\ee
Equation~\eqref{svc0} shows that, in general, the real part of the
HTEE is nonzero. It can nevertheless be set to zero by imposing the
condition $\cos\!\left(\frac{\pi(\theta-1)}{2}\right)=0$, i.e.
\begin{equation}
\theta=-2m,
\qquad m=0,1,2,\dots,
\end{equation}
which yields an infinite family of zeros. We refer to this family as \emph{accidental} zeros, since it occurs only for selected values of the hyperscaling-violation exponent $\theta$ and it can be removed by small deformation of $\theta$ and $\xi$. The most significant case is $\theta=0\ (m=0)$, which corresponds to the AdS$_4$ background whose $\operatorname{Re}(s_{v.c.})$ and $\operatorname{Re}(s_{v.d.})$ solutions were discussed. This well-known background therefore belongs to the accidental zeros, and from this point of view there are infinitely many such zeros. Note that, although $\operatorname{Re}(s_{v.c.})=0$ in this case, the prefactor of the cosine term is nonzero for various values of $\Delta t$ and it diverges as $\Delta t\to0$.

Similarly, \eqref{svd} gives the real part of the $v.d.$ branch as
\begin{equation}\label{svd0}
z_h^{1-\theta}\operatorname{Re}(s_{v.d.})
=\frac{2}{3-\theta}B\!\left(\frac{\theta-1}{3-\theta}.\frac12\right).
\end{equation}

\begin{figure}[htbp]
 \includegraphics[width=0.49\textwidth]{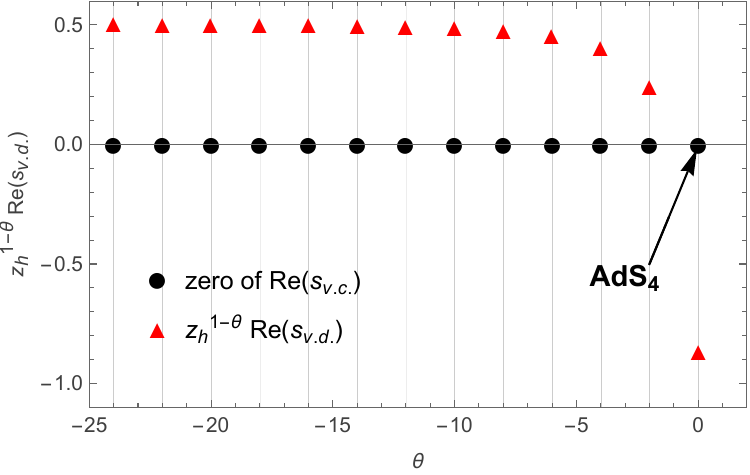}
 \caption{Real part of the $v.d.$ branch (red triangles) and of the 	$v.c.$ branch (black dots) as a
    function of the admissible values of $\theta$, for $d=2$ and
    $\xi=1$. The black dots mark the accidental zeros of
    $\operatorname{Re}(s_{v.c.})$ at $\theta=0,-2,-4,-6,...$. The    	red triangles show $\operatorname{Re}(s_{v.d.})$ evaluated at 		the accidental zeros of the $v.c.$ branch. The AdS$_4$ point $		\theta=0$ is highlighted.}
\label{zeroads4}
\end{figure}

We now consider the cases $m=0,1,2$, or equivalently $\theta=0,-2,-4$. Evaluating the right-hand side of \eqref{svd0} at these points gives
\be
\begin{split}
m=0:\ \frac{2}{3}B\!\left(-\frac{1}{3},\frac12\right)&=-0.86237<0,\\
m=1:\ \frac{2}{5}B\!\left(-\frac{3}{5},\frac12\right)&=0.245273>0,\\
m=2:\ \frac{2}{7}B\!\left(-\frac{5}{7},\frac12\right)&=0.405217>0.
\end{split}
\ee
Except for the \(m=0\) case, all zeros have positive
\(\operatorname{Re}(s_{v.d.})\). Hence the inequality
\begin{equation}\label{condition1}
z_h^{1-\theta}\,\mathrm{Re}(s_{v.c.}) >
z_h^{1-\theta}\,\mathrm{Re}(s_{v.d.})
\end{equation}
does not hold for them, and the \(v.c.\) saddle is favored. The only exception is the AdS\(_4\) solution at \(m=0\), for which
\eqref{condition1} holds. Figure \ref{zeroads4} shows the real parts of both solutions as functions of \(\theta\). The black dots denote the points where \(\operatorname{Re}(s_{v.c.})=0\), while the red triangles show \(\operatorname{Re}(s_{v.d.})\).

The presence of hyperscaling violation and Lifshitz scaling generally breaks the conformal symmetry of the dual field theory. For specific values of these parameters, however, this symmetry is restored. Our results so far show that, in the conformal case, the value of \(\operatorname{Re}(s_{v.d.})\) is negative. As the conformal symmetry is broken further, that is, as \(\theta\) becomes more negative, \(\operatorname{Re}(s_{v.d.})\) increases, and at the other accidental zeros \(\operatorname{Re}(s_{v.d.})\) becomes positive. Thus, the ambiguity in \eqref{condition1} that arises here may be related to this breaking or restoration of conformal symmetry. Moreover, a comparison between the two real parts at values of \(\theta\) other than the zeros is not meaningful. In fact, the zeros occur at isolated points in the
\((\xi,\theta)\) phase space, and altering either parameter makes
\(\operatorname{Re}(s_{v.c.})\) nonzero and dependent on \(\Delta t\), so that our comparison is no longer valid.
\\
\\
\subsection{2+1-Dimensional Case with $\xi=3/2$: General $\theta$ and Accidental Zeros}
For \(d=2\) and \(\xi=3/2\), one has \(D=2-\theta\) and \(\alpha=\theta-5/2\). 
Similar to the previous case, the gravity constraints \eqref{conditions} lead to the following admissible range of \(\theta\)
\be
\theta \in (-\infty,1).
\ee
Then \eqref{svc} and \eqref{svd} reduce to
\be\label{svc-xi32-a}
\operatorname{Re}(s_{v.c.}) =
\frac{B\!\left(\dfrac{\theta-1}{5-2\theta},\dfrac12\right)}{5/2-\theta}
\left[
\dfrac{(5/2-\theta)\,\Delta t}
{B\!\left(\dfrac{\theta-4}{2\theta-5},\dfrac12\right)}
\right]^{2(\theta-1)/3}
\cos\!\left(\dfrac{\pi(\theta-1)}{3}\right),
\ee
and 
\begin{equation}\label{vdtheta}
z_h^{\,1-\theta}\operatorname{Re}(s_{v.d.})
=
\frac{4}{7-2\theta}\,
B\!\left(
\frac{2(\theta-1)}{7-2\theta},
\frac12
\right),
\end{equation}
respectively. Moreover, the real part of the \(v.d.\) solution \eqref{vdtheta} changes sign as 
\[
\setlength{\arraycolsep}{3pt}
\begin{array}{|c|c|c|c|}
\hline
\mathrm{range\ of\ }\theta
& \theta<-\frac{3}{2}
& \theta=-\frac{3}{2}
& -\frac{3}{2}<\theta<1
\\ \hline
\mathrm{Re}(s_{v.d.})
& >0
& =0
& <0
\\ \hline
\end{array}
\]

\begin{figure}[htbp]
 \includegraphics[width=0.5\textwidth]{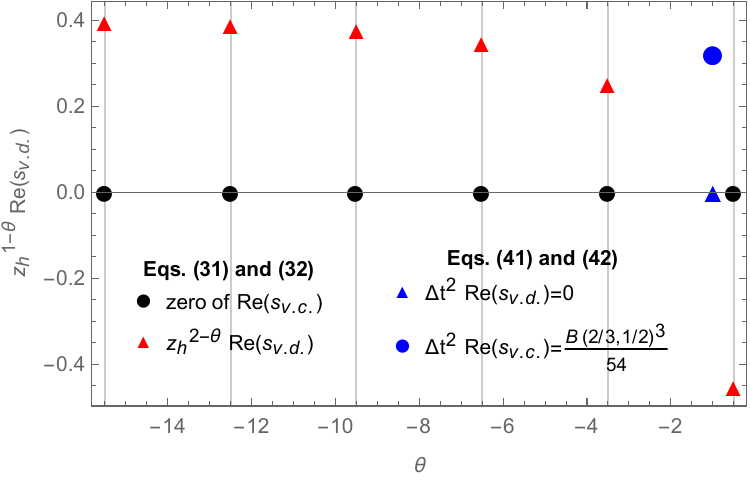}
 \caption{Real part of the \(v.d.\) branch for \(d=2\) and \(\xi=3/2\), plotted as \(z_h^{D-1}\operatorname{Re}(s_{v.d.})\) versus the admissible values of \(\theta\). The black dots mark the accidental zeros of \(\operatorname{Re}(s_{v.c.})\) at \(\theta=5/2-3m\) with \(m=1,2,\dots\), and the red triangles show
\(\operatorname{Re}(s_{v.d.})\) evaluated at these zeros. The blue dot and triangle correspond to \(\operatorname{Re}(s_{v.c.})\) and
\(\operatorname{Re}(s_{v.d.})\), respectively, for \(d=2\), \(\xi=1\), and \(\theta=-1\).}
\label{23}
\end{figure}
The accidental zeros are given by 
\be 
\theta=\frac{5}{2}-3m,\qquad m=1,2,3,\dots
\ee
Similar to the previous case, among the accidental zeros, only the case with \(m=1\) does satisfy \eqref{condition1}, while for all \(m\ge 2\) we have \(\operatorname{Re}(s_{v.d.})>0\), see figure \ref{23} for more illustration.

Therefore, so far the two cases with \(d=2\) and \(\xi=1,\,3/2\) exhibit similar results. In both cases, the first accidental zero satisfies inequality \eqref{condition1}, whereas all other zeros do not. The ambiguity thus arises at this point for \(d=2\), namely for \((\xi=1,\theta=0)\) and \((\xi=3/2,\theta=-1/2)\). The breaking of conformal symmetry does not remove the ambiguity; it persists even in the nonconformal field theory. It is also instructive to note that in both cases \((1-D)/\xi=-1\) and \(\Delta t^{-1}\) appears in the imaginary part of \(s_{v.c.}\). We will return to this point in the next subsection.

\subsection{General 2+1-Dimensional Families: $\Delta t$ Scaling and Criterion Conflict}
In the AdS\(_4\) case, \eqref{vc3} shows that the real part of
\(s_{v.c.}\) vanishes and that its imaginary part is proportional to \(\Delta t^{-1}\). We therefore classify the accidental zeros according to the powers of \(\Delta t\). Since \(d=2\), \(\xi\ge 1\), and \(\alpha<-1\), \eqref{svc} indicates that the scaling of \(s_{v.c.}\) is set by \((i\Delta t)^{(\theta-1)/\xi}\). Different powers of \(\Delta t\) are then obtained by requiring
\be
\frac{\theta-1}{\xi}=-1,-3,\ldots .
\ee
When \(\left|\frac{\theta-1}{\xi}\right|\) is even, the real part of \(s_{v.c.}\) becomes nonzero and depends on the time interval, and therefore is not suitable for comparison with \(\operatorname{Re}(s_{v.d.})\). We will return to this case at the end of the subsection.
Using \eqref{svc}, we obtain
\begin{align}
&\theta=1-\xi:
\qquad s_{v.c.}
=-\,i\,\frac{B\!\left(\frac34,\frac12\right)^2}{4\xi^2}\,\Delta t^{-1},\\
&\theta=1-3\xi:
\qquad
s_{v.c.}=
+\,i\,
\frac{\left|B\!\left(-\frac38,\frac12\right)\right|
B\!\left(\frac58,\frac12\right)^3}{256\,\xi^4}\,
\Delta t^{-3},\\
&\theta=1-5\xi:
\qquad
s_{v.c.}
=
-\,i\,
\frac{\left|B\!\left(-\frac5{12},\frac12\right)\right|
B\!\left(\frac7{12},\frac12\right)^5}{46656\,\xi^6}\,
\Delta t^{-5}.
\end{align}
For $\xi=1$ and \(3/2\), the three cases above correspond precisely to the first three accidental zeros displayed in figures \ref{zeroads4} and \ref{23}. Consequently, for the $v.c.$ solution, the first accidental zero, for any allowed value of $\xi$, is the case in which the imaginary part of $s_{v.c.}$ scales as \(\Delta t^{-1}\), i.e.\ the same power law as in AdS$_4$. The second accidental zero corresponds to \(\Delta t^{-3}\), the third to \(\Delta t^{-5}\), and so on. More generally, for
\be
\theta=1-n\xi,\qquad n=1,3,5,\ldots,
\ee
the real part of $s_{v.d.}$ in \eqref{svd} takes the compact form
\be
z_h^{\,n\xi}\operatorname{Re}(s_{v.d.})
=
\frac{2}{1+(n+1)\xi}\,
B\!\left(-\frac{n\xi}{1+(n+1)\xi},\frac12\right),
\ee
which is independent of $z_h$. Equivalently, for the first few values of $n$,
\begin{align}
\label{acc1}&n=1:\quad
\theta=1-\xi,\quad
z_h^{\xi}\operatorname{Re}(s_{v.d.})
=
\frac{2}{1+2\xi}\,
B\!\left(-\frac{\xi}{1+2\xi},\frac12\right)<0,\\
&n=3:\quad
\theta=1-3\xi,\quad
z_h^{3\xi}\operatorname{Re}(s_{v.d.})
=
\frac{2}{1+4\xi}\,
B\!\left(-\frac{3\xi}{1+4\xi},\frac12\right)>0,\\
&n=5:\quad
\theta=1-5\xi,\quad
z_h^{5\xi}\operatorname{Re}(s_{v.d.})
=
\frac{2}{1+6\xi}\,
B\!\left(-\frac{5\xi}{1+6\xi},\frac12\right)>0.
\end{align}
Note that the sign of $\operatorname{Re}(s_{v.d.})$ alternates with $n$, being negative for $n=1$ and positive for $n=3,5$.
The sign of these beta functions is what decides whether the two selection criteria agree or disagree. At an accidental zero the real part of the $v.c.$ solution vanishes, so the minimal-real-part criterion selects the $v.d.$ branch whenever $\operatorname{Re}(s_{v.d.})<0$, and the $v.c.$ branch whenever $\operatorname{Re}(s_{v.d.})>0$. The vacuum-reduction criterion, on the other hand, always selects the $v.c.$ branch, since only that branch reduces to the vacuum answer as $\Delta t\to 0$. Hence the two criteria conflict if and only if $\operatorname{Re}(s_{v.d.})<0$. The above equations show that this happens only for $n=1$; for $n=3,5,\ldots$ the real part of $s_{v.d.}$ is positive and both criteria select the $v.c.$ branch, so no ambiguity arises.
Thus the ambiguity appears only at the first accidental zero, where the real part of $s_{v.d.}$ is negative and $s_{v.c.}$ scales as \(\Delta t^{-1}\). In the AdS$_4$ case, this scaling is a consequence of the conformal symmetry of the field theory. In the presence of hyperscaling violation and/or Lifshitz scaling, conformal symmetry is broken; nevertheless, for suitable values of \(\theta\) and \(\xi\), the same \(\Delta t\)-power law as in the conformal vacuum appears at the first accidental zero. We refer to it as the closest analogue of the AdS$_4$ point in the \(\xi\) family: it is the unique accidental zero at which the $v.c.$ branch retains the conformal $\Delta t^{-1}$ scaling. It is precisely at this point that the two criteria disagree, and the ambiguity therefore persists there.

At the first accidental zero, \(\theta=1-\xi\), the real part of the $v.d.$ branch is given by \eqref{acc1}. For every admissible finite \(\xi\ge 1\), the argument of the beta function satisfies
\be
-\frac{\xi}{1+2\xi}\in\left(-\frac12,-\frac13\right],
\ee
and in this interval \(B(x,\tfrac12)\) is strictly negative and nowhere vanishing. Consequently, for all finite $\xi\ge 1$, we have ${\rm Re}(s_{v.d.})<0$ and as a result there is no physical value of \(\xi\) for which \(\operatorname{Re}(s_{v.d.})=0\). The beta function vanishes only when
\be
-\frac{\xi}{1+2\xi}=-\frac{2k+1}{2},
\qquad k=1,2,\dots,
\ee
that is, at
\be
\xi=-\frac{2k+1}{4k}<0,
\ee
which lies outside the admissible range \(\xi\ge 1\). The only way to approach \(\operatorname{Re}(s_{v.d.})=0\) is the unattained limit \(\xi\to\infty\), where the beta-function argument approaches \(-1/2\) from above and the beta function tends to zero from below. Thus the conflict between the minimal-real-part criterion and the vacuum-reduction criterion at the first accidental zero is not tied to any special value of \(\xi\); it persists for all admissible \(\xi\) and disappears only in the infinite-Lifshitz-exponent limit.

Another interesting possibility is \(d=2\) with \(\theta=1-2\xi\) $(n=2)$. In this case, \eqref{svc} and \eqref{svd} give
\be
\begin{split}
s_{v.c.}&=\frac{B\!\left(\frac{2}{3},\frac{1}{2}\right)^3}{54\,\xi^3\,\Delta t^2},\\
\operatorname{Re}(s_{v.d.})&=\frac{2z_h^{-2\xi}}{1+3\xi}\,
B\!\left(-\frac{2\xi}{1+3\xi},\frac12\right).\\
\end{split}
\ee
Setting the real part of the \(v.d.\) solution to zero, we find that for \(\xi=1\) the beta function \(B(-1/2,1/2)\) vanishes independent of $z_h$. Consequently, at \(\xi=1\),
\be\label{newcondition}
\Delta t^{2}\operatorname{Re}(s_{v.c.})
>
\Delta t^{2}\operatorname{Re}(s_{v.d.}).
\ee
Since \(s_{v.c.}\) is real and positive for \(\Delta t>0\), a new
ambiguity of the same type appears, now between the solution selected by the minimal-real-part criterion and the one selected by the vacuum-reduction criterion. The blue dot and triangle in
figure \ref{23} show \(\Delta t^{2}\operatorname{Re}(s_{v.c.})\) and \(\Delta t^{2}\operatorname{Re}(s_{v.d.})\), respectively. The case \(n=2\) is the only one that leads to \eqref{newcondition}; all other even values of \(n\) do not satisfy this inequality.

\section{Vacuum-connected and disconnected HTEE in 3+1 field theories}
We investigate the behavior of the $v.c$ and $v.d.$ solutions in $3+1$-dimensional field theories in this section. Based on our results in the previous section, we consider the general case $d=3$, for which the power of \(i\Delta t\) in \eqref{svc} takes various admissible values. We therefore set
\be
\frac{\theta-2}{\xi}=-1,-3,-5,\ldots,
\ee
and \eqref{svc} then gives
\bse\label{vc33}\begin{align}
\theta=2-\xi:\quad & s_{v.c.}=-\,i\,\frac{B\!\left(\frac34,\frac12\right)^2}{4\xi^2}\,\Delta t^{-1},\\[8pt]
\theta=2-3\xi:\quad & s_{v.c.}=+\,i\,\frac{\left|B\!\left(-\frac38,\frac12\right)\right|B\!\left(\frac58,\frac12\right)^3}{256\,\xi^4}\,\Delta t^{-3},\\[8pt]
\theta=2-5\xi:\quad & s_{v.c.}=-\,i\,\frac{\left|B\!\left(-\frac5{12},\frac12\right)\right|B\!\left(\frac7{12},\frac12\right)^5}{46656\,\xi^6}\,\Delta t^{-5}.
\end{align}\ese
Defining
\be
\theta=2-n\xi,\qquad n=1,3,5,\ldots,
\ee
\eqref{svd} becomes
\be
z_h^{\,n\xi}\,\mathrm{Re}\!\left(s_{v.d.}\right)=
\frac{2}{1+(n+1)\xi}\,B\!\left(-\frac{n\xi}{1+(n+1)\xi},\frac12\right).
\ee
Equivalently, we obtain
\bse\label{vd33}\begin{align}
n=1:\quad \theta=2-\xi,\ \ & z_h^{\,\xi}\,\mathrm{Re}\!\left(s_{v.d.}\right)=\dfrac{2}{1+2\xi}\,
B\!\left(-\dfrac{\xi}{1+2\xi},\frac12\right)<0,\\[10pt]
n=3:\quad \theta=2-3\xi,\ \ & z_h^{\,3\xi}\,\mathrm{Re}\!\left(s_{v.d.}\right)
=\dfrac{2}{1+4\xi}\,
B\!\left(-\dfrac{3\xi}{1+4\xi},\frac12\right)>0,\\[10pt]
n=5:\quad \theta=2-5\xi,\ \ & z_h^{\,5\xi}\,\mathrm{Re}\!\left(s_{v.d.}\right)
=\dfrac{2}{1+6\xi}\,
B\!\left(-\dfrac{5\xi}{1+6\xi},\frac12\right)>0.
\end{align}
\ese
Comparing \eqref{vc33} and \eqref{vd33}, we see that the ambiguity 
\begin{equation}\label{condition111}
z_h^{n\xi}\,\mathrm{Re}(s_{v.c.}) >
z_h^{n\xi}\,\mathrm{Re}(s_{v.d.}),
\end{equation}
 arises for the first equation, $n=1$, namely the \(\theta=2-\xi\) family, as in the \(d=2\) case with \(\theta=1-\xi\). For the other families, however, the \(v.c.\) solution is favored. It is important to note that the same power of \(\Delta t\), i.e.\ \(\Delta t^{-1}\), is the origin of the ambiguity in both \(d=2\) and \(d=3\). 

In the special case with $d=3$, $\xi=1$ and $\theta=0$, the geometry reduces to AdS$_5$ and the $v.c.$ solution gives
\be\begin{split} 
s_{v.c.}&=\frac{B\!\left(\frac{2}{3},\frac{1}{2}\right)^3}{54\,\Delta t^2}>0,\cr
s_{v.d.}&=iz_h^{-2},
\end{split}\ee
and it can be seen the $s_{v.c.}$ is real and proportional to $\Delta t^{-2}$. This result is supported by conformal symmetry as $\Delta t^{-1}$ in $d=2$ dimension. However, $s_{v.d.}$ is purely imaginary and the real part of f $s_{v.d.}$ saddle is thus zero. 
Therefore, in the AdS$_{5}$ background there is also ambiguity and the inequality \eqref{newcondition} persists in $3+1$ dimensional confromal field theory, that the minimal-real-part criterion selects the $v.d.$ saddle, while the vacuum-reduction criterion selects the $v.c.$ branch.

\section{Discussion: Negative and Positive Ambiguity Families}
We have analyzed the saddle-selection ambiguity in the HTEE for
hyperscaling-violating Lifshitz backgrounds. The two natural criteria are the minimal-real-part criterion, which compares the on-shell actions of \(v.c.\) and \(v.d.\)
complex extremal surfaces, and the vacuum-reduction criterion, which requires the physical saddle to reduce to the vacuum answer as \(\Delta t\to0\). For \(d=2\) with \(\xi=1\) and \(\xi=3/2\), as well as for the general families \(\theta=1-n\xi\) with \(n=1,3,5,\ldots\), the two criteria agree everywhere except at the first accidental zero, where \(\mathrm{Re}(s_{v.c.})=0\) and \(\mathrm{Re}(s_{v.d.})<0\). The same structure appears in \(d=3\) for the family \(\theta=2-n\xi\): only \(n=1\) is ambiguous, while for \(n=3,5,\ldots\) the real part of \(s_{v.d.}\) is positive and both criteria select the \(v.c.\) branch.

As a final step, we would like to determine whether the scaling
\(\Delta t^{-1}\) of the imaginary part of \(s_{v.c.}\) exists in general
\(d>2\) and leads to an ambiguity. We have observed that this is the case
for \(d=2\) and \(d=3\). Therefore, for hyperscaling-violating Lifshitz
backgrounds in general \({d+2}\) dimensions, the conditions in
\eqref{conditions} are satisfied for \(\frac{1-D}{\xi}=-1\) if
\be\label{aa}
\theta=d-1-\xi, \qquad
\xi\ge \max\left\{1,\frac{2d-1}{d+1}\right\}.
\ee
Thus \eqref{svc} and \eqref{svd} lead to
\be
\operatorname{Re}(s_{v.c.})=0,
\qquad
\operatorname{Re}(s_{v.d.})= \frac{2z_h^{-\xi}}{1+2\xi} B\!\left(
-\frac{\xi}{1+2\xi}, \frac12 \right)<0.
\ee
This shows that the condition \eqref{condition} is generalized to
\be
z_h^\xi ~ {\rm Re}(s_{v.c.}) > z_h^\xi ~ {\rm Re}(s_{v.d.}).
\ee
As a result, in general \(d\) dimensions, there exist values of
\(\xi\) and \(\theta\), and corresponding backgrounds, for which
\eqref{svc} leads to \((i \Delta t)^{-1}\) and the ambiguity persists.

According to our results, the ambiguity appears in two distinct classes.
\begin{itemize}
\item \textit{Negative family}: In this class, the real parts of the
\(v.c.\) and \(v.d.\) solutions are zero and negative, respectively, so
that
\begin{equation}\label{nn}
z_h^{D-1}\,\operatorname{Re}(s_{v.c.}) >
z_h^{D-1}\,\operatorname{Re}(s_{v.d.}),
\end{equation}
holds. For admissible backgrounds specified by suitable values of
\(\xi\) and \(\theta\), the imaginary part of \(s_{v.c.}\) scales as
\(\Delta t^{-1}\), as in the \(2+1\)-dimensional conformal field theory.
Surprisingly, the same structure appears in the \(3+1\)-dimensional
nonconformal field theory. It therefore seems that the origin of the
ambiguity \eqref{nn} is the \(\Delta t^{-1}\) scaling in these field
theories. In summary, we have \({\rm Re}(s_{v.c.})=0\), and \eqref{nn}
holds for
\begin{equation}
{\rm Im}(s_{v.c.})\propto \Delta t^{-1}:
\begin{cases}
\text{2+1-dimensional conformal field theory}\ (d=2,\ \xi=1,\ \theta=0),\\
\text{2+1-dimensional nonconformal field theory}\ (d=2,\ \xi\ge 1,\ \theta=1-\xi),\\
\text{3+1-dimensional nonconformal field theory}\ (d=3,\ \xi\ge 5/4,\ \theta=2-\xi),\\
\text{$d$+1-dimensional nonconformal field theory }(d,\ (\xi,\theta)\ \text{as in \eqref{aa}}).
\end{cases}
\end{equation}

\item \textit{Positive family}: In contrast to the previous case, the real
part of the \(v.c.\) solution is positive and that of the \(v.d.\)
solution vanishes, so that
\begin{equation}\label{mm}
\Delta t^{2}\operatorname{Re}(s_{v.c.})>
\Delta t^{2}\operatorname{Re}(s_{v.d.}),
\end{equation}
holds. This family arises in the conformal field theory in \(3+1\)
dimensions as well as in a nonconformal field theory in \(2+1\)
dimensions with the specific values \(\xi=1\) and \(\theta=-1\). This
power-law scaling, i.e. \(\Delta t^{-2}\), is protected by conformal
symmetry in the \(3+1\)-dimensional conformal field theory. In summary,
\({\rm Re}(s_{v.d.})=0\), and \eqref{mm} holds for
\begin{equation}
{\rm Re}(s_{v.c.})\propto \Delta t^{-2}:
\begin{cases}
\text{2+1-dimensional nonconformal field theory}\ (d=2,\ \xi=1,\ \theta=-1),\\
\text{3+1-dimensional conformal field theory}\ (d=3,\ \xi=1,\ \theta=0).
\end{cases}
\end{equation}
\end{itemize}
As a result, one can conclude that the ambiguity is not tied to the
breaking or restoration of conformal symmetry.


\appendix
\section{Momentum Scaling in the $v.c.$ and $v.d.$ Limits}\label{A1}
The conditions \eqref{alphacondition}, considered in \cite{Heller:2024whi}, depend in our case on the value of \(\alpha=1-\xi-D\). For general \(\alpha\), the \(v.c.\) and \(v.d.\) solutions can be classified according to the behavior of \(|p|\) in the limits \(z_*\to0\) and \(z_*\to\infty\), as summarized in the table below:
\[
\begin{array}{|c|c|c|}
\hline
 & z_*\to 0\ (\text{v.c.}) & z_*\to \infty\ (\text{v.d.})\\[2mm]
\hline
|p|\ \text{behavior} 
& |p|\sim z_*^{\alpha} 
& |p|\sim z_*^{(1+\alpha)/2} \\[1mm]
\hline
\alpha<-1 & \infty & 0 \\
\alpha=-1 & \infty & \text{finite} \\
-1<\alpha<0 & \infty & \infty \\
\alpha=0 & \text{finite} & \infty \\
\alpha>0 & 0 & \infty\\
\hline
\end{array}
\]
The AdS$_4$ and AdS$_5$ cases lie in the \(\alpha<-1\) regime, so we also focus on this limit. Other values of \(\alpha\) are likewise worth discussing; we defer this to future work.

\end{document}